\documentclass[journal=JPCLett,manuscript=article]{achemso}

\usepackage[version=4]{mhchem}
\usepackage{graphicx}
\usepackage{amssymb,amsmath}
\usepackage{color}
\usepackage{calc}
\usepackage{xspace}
\usepackage{soul} 
\usepackage{hyphenat}
\usepackage{braket}

\author{Joseph Holmes}
\affiliation[Department of Chemistry]
{Department of Chemistry, Indiana University, Bloomington, IN 47405, U.S.}
\altaffiliation{Contributed equally to this work}

\author{Arathi Anil Sushma}
\affiliation[Department of Chemistry]
{Department of Chemistry, Indiana University, Bloomington, IN 47405, U.S.}
\altaffiliation{Contributed equally to this work}

\author{Irina B. Tsvetkova}
\affiliation[Department of Chemistry]
{Department of Chemistry, Indiana University, Bloomington, IN 47405, U.S.}
\email{itsvetko@indiana.edu}

\author{William L. Schaich}
\affiliation[Physics Department]
{Physics Department, Indiana University, Bloomington, IN 47405, U.S.}

\author{Richard D. Schaller}
\affiliation[CNM]
{The Center for Nanoscale Materials at Argonne National Laboratory, Lemont, IL 60439, U.S.}

\author{Bogdan Dragnea}
\affiliation[Department of Chemistry]
{Department of Chemistry, Indiana University, Bloomington, IN 47405, U.S.}
\email{dragnea@indiana.edu}

\title[]
  {Ultrafast Collective Excited State Dynamics of a Virus-supported Fluorophore Antenna}


\begin{document}

\begin{tocentry}

    \centering
    \includegraphics[width = 1 \textwidth]{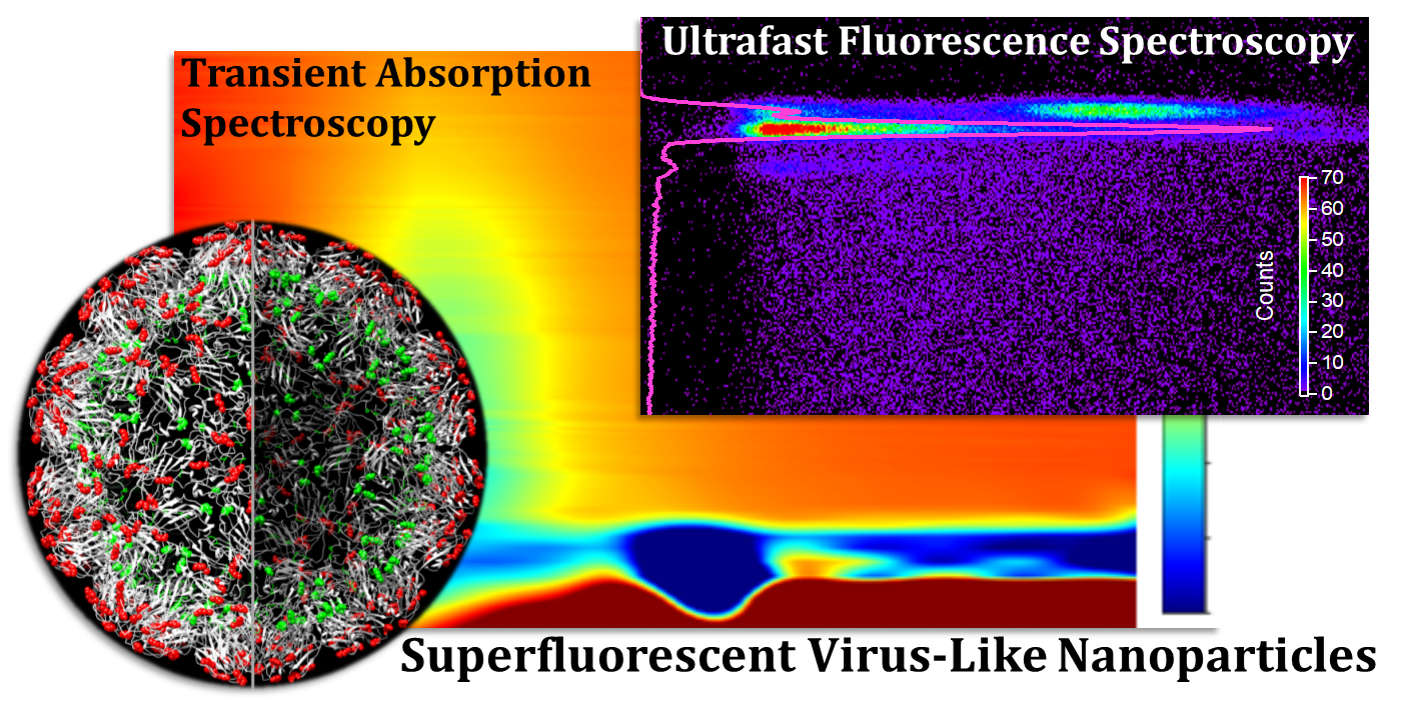}

\end{tocentry}

\begin{abstract}

Radiation brightening  was recently observed in a multi-fluorophore-conjugated brome mosaic virus (BMV) particle, at room temperature under pulsed excitation. Based on its nonlinear dependence on the number of fluorophores,  the origins of the phenomenon were attributed to a collective relaxation. However, the mechanism remains unknown. We present ultrafast transient absorption and fluorescence spectroscopic studies which shed new light on the collective nature of the relaxation dynamics in such radiation-brightened, multi-fluorophore particles. Our findings indicate that the emission dynamics is consistent with a superradiance mechanism. The ratio between the rates of competing radiative and non-radiative relaxation pathways depends on the number of fluorophores per virus. We also discuss the evidence of coherent oscillations in the transient absorption trace from multi-fluorophore conjugated which last for $\sim100$s of picoseconds, at room temperature. The findings suggest that small icosahedral virus shells provide a unique biological scaffold for developing non-classical, deep subwavelength light sources, and may open new realms for the development of photonic probes for medical imaging applications.
  
\end{abstract}


High-contrast luminescent nanoprobes enable a myriad applications including  biological detection\cite{Wolfbeis2015}, therapeutics\cite{yi2014near,vats2017near}, sensing\cite{ueno2011fluorescent,gao2021fluorescent}, optogenetics\cite{leopold2019fluorescent,oheim2014new}, and anti-counterfeiting\cite{abdollahi2020photoluminescent,li2018tunable}. For the vast majority of current probes, radiation is the result of random, spontaneous relaxation. Consequently, emission dynamics obeys the classical exponential decay \cite{Valeur2013}. Since background emission has similar dynamics, time-domain background removal to improve contrast is seldom a viable option. Increasing the number of emitters per nanoprobe to augment brightness generally results in self-quenching due to inter-emitter distances becoming short enough ($ \lesssim 5$ nm) for efficient resonant energy transfer to occur. Both challenges could be addressed by constructing a multi-emitter nanoprobe with correlated, non-classical emission. In this letter we present experimental evidence for this behavior, which occurs after pulsed excitation of a dense array of hundreds of fluorescent dyes, deterministically-arranged on a 28 nm diameter icosahedral virus template.

Instead of the several ns exponential decay expected from individual fluorophores in solution, emission from a multi-fluorophore virus particle,  at saturation coverage, occurs as a short burst of $\approx 20 $ ps. Peak intensity is attained at $\approx 25$ ps after the ultrafast excitation pulse. Instead of being nearly quenched like under cw excitation, the estimated quantum yield in burst-mode is comparable to that of free, individual fluorophores. These unusual characteristics occur at room temperature, in a biocompatible setting. Therefore, such viromimetic probes are promising to overcome some of the current limitations of classical biophotonic probes and open new venues in fluorescence imaging. 

Radiation brightening from dye-conjugated fluorescent virus-like particles (fVLPs) was first reported  by Tsvetkova \emph{et al} \cite{tsvetkova2019radiation}. It was found that when the fluorophores are conjugated with reactive residues of the brome mosaic virus (BMV) capsid interfaces, emission by the complex is strongly accelerated with respect to that of the free dye \cite{tsvetkova2019radiation}. The brightening effect was found to be a nonlinear function of $N$, the average number of fluorophores per particle. Thus, the origins of the phenomenon were attributed to a collective effect \cite{tsvetkova2019radiation, AnilSushma2021}. However, emission dynamics, which carries potential clues about the mechanism, remained unknown. To obtain additional information about the mechanism of radiation brightening in fVLPs, we performed measurements of the emission and the excited state relaxation dynamics.

\begin{figure}[t]
    \centering
    \includegraphics[width = 1 \textwidth]{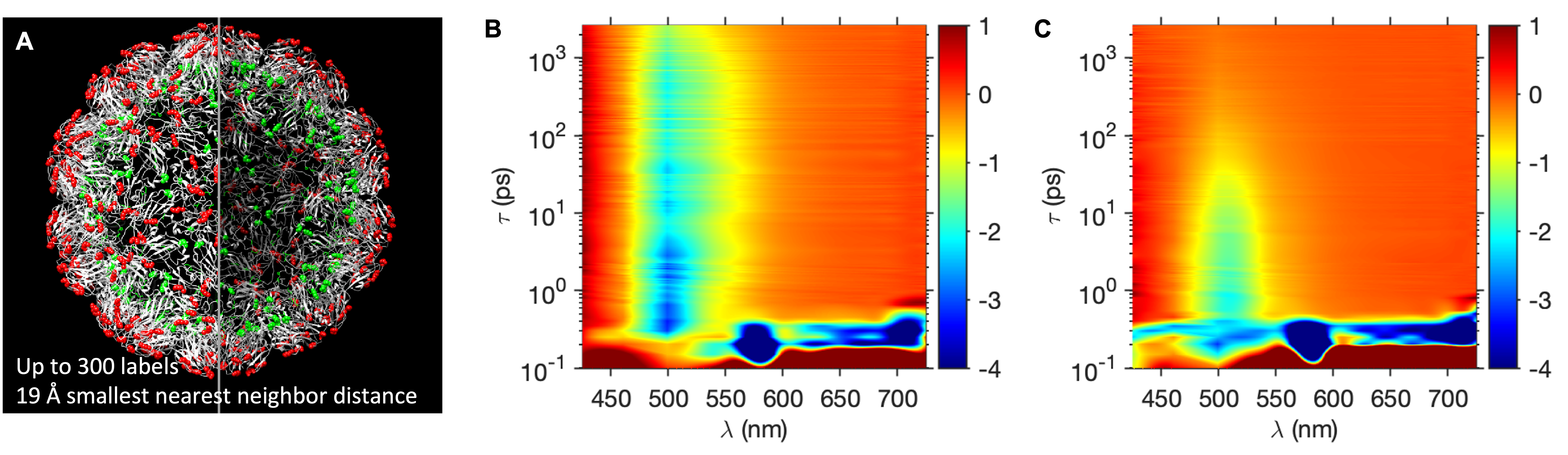}
    \caption{A) Schematic of accessible labelling sites on BMV capsid: red – external lysines K64,K105, K165 and green – internal lysines K41, K44 and K86. B) and C) transient absorption surface plots of free OG and BMV-OG278 following an initial excitation at 7.1 $\text{mJ}/\text{cm}^2$. The color scale corresponds to the difference between the absorbance of the samples in excited and ground-state (units of m$\Delta A$)}
    \label{scheme}
\end{figure}

 In this work, a fluorescein-derived dye, Oregon Green\textsuperscript{TM} 488, was covalently bound to the BMV capsid \emph{via} NHS ester labelling of exposed lysines\cite{tsvetkova2019radiation}. Maximum labeling density was $\sim300$ dyes/virus\cite{tsvetkova2019radiation}(see Figure SI1). Figure~\ref{scheme}A shows a molecular model of BMV, with the dye-accessible external and internal lysines colored in red and green respectively\cite{AnilSushma2021}. Here "internal" means located between the lumenal or outer capsid surfaces. The minimum nearest neighbour distance between lysines was estimated to be $1.9~$nm\cite{tsvetkova2019radiation}. 
 
To follow the excited state dynamics of coupled fluorophores in fVLPs, we performed pump-probe femtosecond, transient absorption (TA) spectroscopy. Differential absorption spectra of isolated fluorophores and fVLP samples with an average number of 278 dyes per particle (hereby called BMV-OG278) are presented as a function of the wavelength and time-delay in Figure~\ref{scheme}B and Figure~\ref{scheme}C, respectively. The ultrafast excitation pulse at 488 nm was provided by an optical parametric amplifier pumped by a regenerative amplifier ($\sim 120$ fs pulse width).

In both the control and BMV-OG278 sample, two intense negative signals are evident at very early times in the transient spectra. One appears at $570-580$ nm and it can be attributed to Raman scattering of water\cite{parker1959raman}. Since this event is simultaneous with the excitation pulse, its appearance defines the zero delay between pump and probe pulses. The second, early negative signal is due to the coherent interaction between pump and probe pulses, which leads to stimulated Raman amplification appearing at 650-750 nm\cite{Lorenc2002}. Both Raman signals vanish within $0.4$ ps from the pump pulse. 

A spectral region of particular interest is $500-530$ nm, where ground state bleach (GSB) is expected. Indeed, both sample and control exhibit a prominent feature in this spectral region. However, there are also some stark differences between the two, Figure~\ref{scheme}B and Figure~\ref{scheme}C: In particular, the excited state decay is much faster for the BMV-OG278 sample than for the free dye solution.

Minima in $\Delta A$ spectra were at $498$ nm for free dye and $504$ nm for VLP-bound fluorophores, and are consistent with the peak wavelength in steady-state absorption spectra (Figure~SI1). The TA spectral evolution of the free dye solution and BMV-OG278 samples depends on pump energy, Figures~ SI3-7. The $5-6$ nm shift between free and bound fluorophores is attributable to the difference in dielectric constant between protein and water.  

\begin{figure}[ht]
    \centering
    \includegraphics[width = 0.75\textwidth]{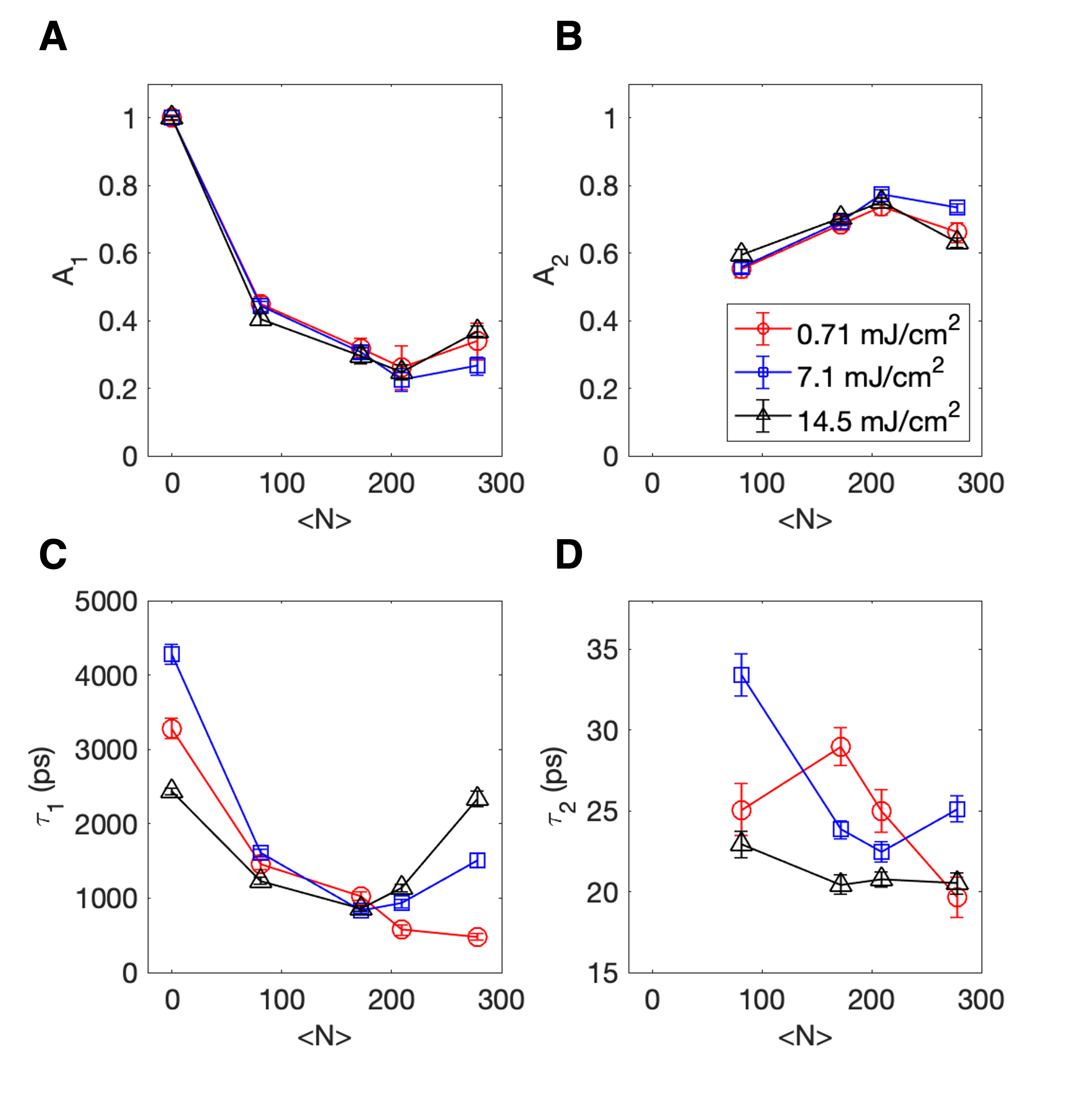}
    \caption{Fit components for the excited state decay dynamics plotted against varying $\braket{N}$ \emph{i.e.} number of fluorophores/virus; Top panel: amplitudes $A_1$ and $A_2$; Bottom panel: Lifetimes ($\tau_1$) and ($\tau_2$)  }
    \label{TA}
\end{figure}

For quantitative time-domain analysis and comparison, TA spectra were integrated from 520 to 530 nm and fitted with an exponential model from $400$ fs to $2.6$ ns (Figure SI8-12). For analysis, the wavelength and time-delay range were chosen past the ground state bleach minimum to avoid artifacts from pump laser scattering as well to remove coherent artifacts which arise around the zero delay time where the pump and probe beams are temporally overlapping\cite{vardeny1981picosecond,hui2011elimination}. Exponential fit parameters (amplitude and decay time) were obtained by means of nonlinear least squares, and the estimates of the errors from the model were computed from the sample covariance matrix. Best fit kinetic models for the transients were found to be predominantly mono-exponential for the free fluorophore and bi-exponential for the virus-bound fluorophore. Normalized amplitudes ($A_i$) and decay times ($\tau_i$) obtained from the fit procedure are presented as a function of $\braket{N}$ in the plots in Figure~\ref{TA}, with $\braket{N}=0$ corresponding to free fluorophores in solution. 

At low power, the excited state lifetime ($\tau_1$) of free dyes measured by TA was in average $\approx 3.3$ ns, which matches the spontaneous emission lifetime of OG in solution. Since the quantum yield is high for this dye, fluorescence emission is the dominant pathway in the relaxation dynamics of the free fluorophore excited at $488~$nm. The long lifetime component ($\tau_1$) exhibits a monotonic shortening as the number of bound dyes increases (Figure~\ref{TA}C, red curve). At high labeling density, the $\tau_1$ is reduced by a factor of $6$ when compared to the free fluorophore. In previous fluorescence lifetime experiments at similar excitation fluence and $\braket{N}$ the fluorescence lifetime was short, and the photon counts were quenched \cite{tsvetkova2019radiation}. A possible explanation is that at low pump power, only a few fluorophores are excited while the majority are in the ground state and there is a high probability for homo- resonance energy transfer (RET) to occur, with the result of an increase in fluorescence quenching. In any event, we note a correlation between the evolution of the TA $\tau_1$ evolution and that of the fluorescence decay measured previously.

As the pump fluence increases, $\tau_1$ decreases for the free dye, possibly due to photobleaching  or  to an increase  in inter-system crossing and triplet state formation \cite{Demchenko2020}. Indeed, some photobleaching was observed in amount of $10-20\%$ between the first and second run for each sample. However, at high labeling density ($\braket{N} \gtrsim 200$) $\tau_1$ increases from $\approx 600$ ps at lowest pump fluence up to $\approx 2000$ ps for highest pump fluence. Thus, it appears that the nonradiative rate, which was deemed responsible for the initial shortening of the lifetime at low fluences and high density \cite{tsvetkova2019radiation}, slows down to a value that is observed at lower labeling densities. 

The second fit component (parameters, $A_2$ and $\tau_2$) dominates BMV-OG dynamics at short times suggesting a new relaxation channel that only operates in the multi-fluorophore VLP. The associated amplitude $A_2$ is significantly greater than $A_1$, Figure~\ref{TA}B. Its time constant is $\tau_2\approx 25-35$ ps for low and medium fluence and $\approx 20-25$ ps for the highest pump fluence, Figure~\ref{TA}D. This component appears to decrease with the number of dyes per virus although the trends are noisy (Figure~\ref{TA}D).

To summarize up to this point, two factors appear to lead to changes in the excited state dynamics: i) pump fluence at fixed $\braket{N}$, and ii) $\braket{N}$ at all fluences. Specifically, relaxation dynamics is accelerated when $\braket{N}$ increases. 

\begin{figure}[t]
    \centering
    \includegraphics[width =1 \textwidth]{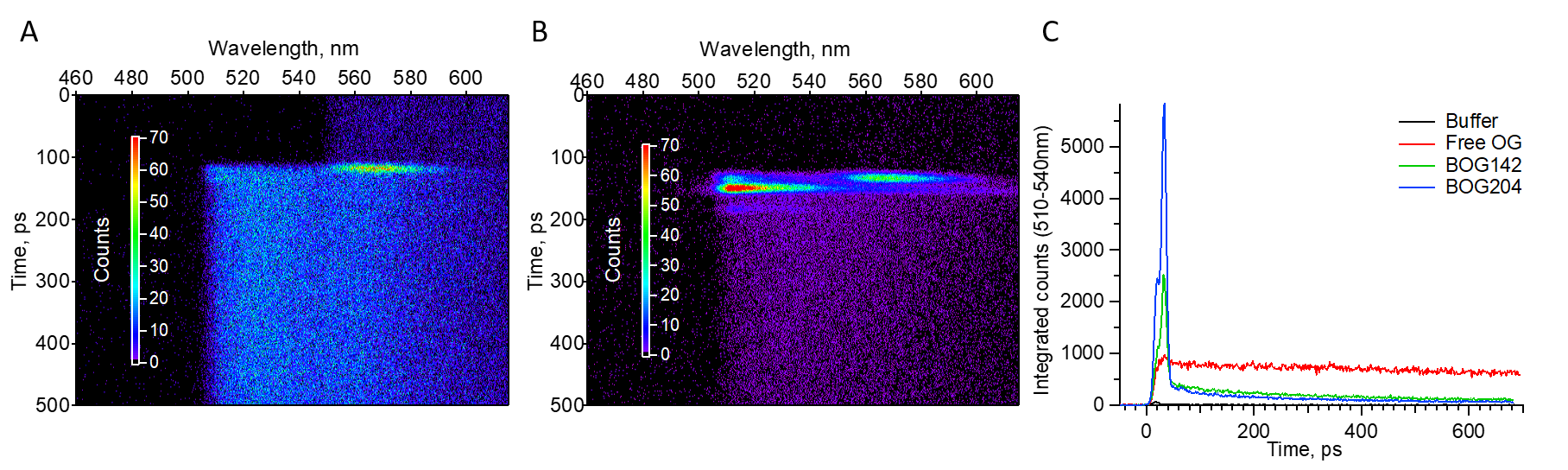}
    \caption{Dynamics of fluorescence emission using a streak-camera detector.The total dye concentration for all samples was kept at 200 nM, and the laser energy density was 16.2 mJ/cm$^\mathrm{2}$. A) Free OG dye. B) BMV-OG204. ; C) Wavelength-integrated (510-540 nm), time-resolved fluorescence decay traces for BMV-OG and controls.}
    \label{streak}
\end{figure}

While fluorophore coupling clearly affects relaxation lifetimes, the result of the competition between nonradiative and radiative relaxation channels cannot be solely gauged based on TA. To obtain further spectral and temporal information on fluorescence emission, we have performed time-resolved fluorescence spectroscopy with a streak-camera detector. Figures \ref{streak}A and B present spectrally and time-resolved fluorescence emission for free fluorophore and BMV-OG204, respectively. The pump pulse arrives at $t_0 \sim 110$ ps. Raman scattering from water, at $\approx560-580$ nm, is simultaneous with the pump pulse. Also, a faint spot at $t_0$ can be observed at $\approx500-510$ nm in free OG and BMV-OG samples. This is due to elastic scattering of the laser line being incompletely suppressed by laser rejection filters. The scattering peak is very faint in the case of free OG because this sample does not scatter as strongly as the virus particles. The inelastic and the elastic scattering peaks are convenient because they facilitate the timing of fluorescence emission with respect to the pump pulse. 

For free dye samples emission is spread in time and decays uniformly along the entire measured time interval, Figure \ref{streak} A,C. In stark contrast, the fluorescence from the BMV-OG samples is emitted as a burst of $\approx20$ ps duration. The burst is delayed with respect to the pump pulse, with a maximum occurring at $t \approx25$ ps after excitation, Figure \ref{streak}B,C. At this point, we note that the value of $\tau_2$ from TA measurements is close to the characteristic time of the fluorescence burst observed in the streak camera experiment. This observation is consistent with the short relaxation pathway observed in TA corresponding to radiative relaxation. The wavelength-integrated peak amplitude of the burst shows a $2 \times$ increase when $\braket{N}$ is increased from $\approx 140$ to $\approx200$ dyes per virus, Figure \ref{streak}C.

What phenomena could potentially be responsible for delayed, non-exponential, accelerated radiation in VLPs? We consider superfluorescence (SF), superradiance (SR), and amplified spontaneous emission (ASE)\cite{Cong2016,dai2011brief}. Although we are not completely excluding the possibility of lasing, i.e. light amplification by stimulated emission, we note that it is less likely, at least in the conventional sense, because of the absence of a resonator and the extremely small VLP volume and thus, small Purcell factor\cite{Samuel2009}. 

There are qualitative differences between ASE and SF in both time and frequency domain. In the frequency domain, SF(SR) emission occurs in a broader window\cite{Dicke1954} while in ASE, spectral narrowing is expected as pump intensity increases\cite{allen1973amplified}. We have not observed spectral narrowing in our experiments\cite{tsvetkova2019radiation}.

In the time domain, the SF pulse is sharp and develops after a time delay \cite{Malcuit1987}. In ASE, the time delay is vanishingly small, the output pulse is longer than in SF and noisy, and the pulse duration is insensitive to dephasing factors (e.g. local viscosity and temperature)\cite{wilk1983laser,rai1992quantum,maki1989influence}. From previous work \cite{tsvetkova2019radiation, AnilSushma2021} we know that the radiation brightening from multi-fluorophore VLPs is very sensitive to the local environment, unlike ASE.  Therefore, the significant shortening of excited state lifetime observed in TA and streak camera experiments, the delayed intense pulse in fluorescence emission, the broad fluorescence spectrum, the sensitivity to the fluorophore local environment, all point to a superradiance-like mechanism responsible for radiation brightening. We discuss a simple superradiance model in the SI, which illustrates the burst emission. 

An intriguing feature here is the fact that the phenomenon occurs at room temperature. SR emission dynamics was experimentally studied first in atomic gases at very low temperatures \cite{Gross1982a}. Later, it was detected in low-temperature solids\cite{miyajima2017ultrashort,tighineanu2016single,raino2018superfluorescence,krieg2020monodisperse,haider2021superradiant}. Excitonic superradiance was studied extensively in molecular aggregates, at cryogenic temperatures as well. \cite{Palacios2002,engel2007evidence}. 

Only recently SR was shown to occur at room temperature in photosynthetic complexes \cite{Monshouwer1997,Malina2021}. The structured environment within the photosynthetic complex is believed to promote collective fluorophore behavior since rigidity and proper orientation can alleviate thermal dephasing  \cite{rolczynski2018correlated}. In support of this idea, alterations in the relative position and orientation of fluorophores in the protein pockets of an artificial photosynthetic system were found to strongly affect the excited state dynamics of protein-bound fluorophores \cite{noriega2015manipulating,delor2018exploiting}. However, it is worth emphasizing that interfluorophore interactions in light harvesting complexes and in most molecular J-aggregates are characterized by strong coupling \cite{Spano1989, meinardi2003superradiance,ratner2003superradiance}. By comparison, the nearest-neighbor fluorophore distances on BMV-OG are much longer than that of molecular aggregates, which precludes electron tunnelling. 

In our case, collective emission of radiation arise from weak coupling at room temperature. Until now, this coupling range has not been given much attention. Dipole-dipole coupling is considered detrimental to the formation of coherent multi-fluorophore states \cite{Gross1982a}. Could the virus template play a role in coupling? The original model of Dicke superradiance (see SI) does not take into account the spatial organization or orientation of the fluorophores. However, from previous work it was clear that the nature of the template is important for the radiation brightening effect \cite{tsvetkova2019radiation,AnilSushma2021}. Below we provide further evidence for it from a study of relaxation dynamics.

We performed additional TA experiments with a set-up in which pump and probe beams were counter-propagating, at a small angle. 
This set-up change allowed us to reduce laser scattering artifacts and measure the excited state dynamics at high laser powers with the probe wavelength closer to ground state bleach peak position $\lambda=500-520$ nm. 

\begin{figure}[t]
    \centering
    \includegraphics[width =1 \textwidth]{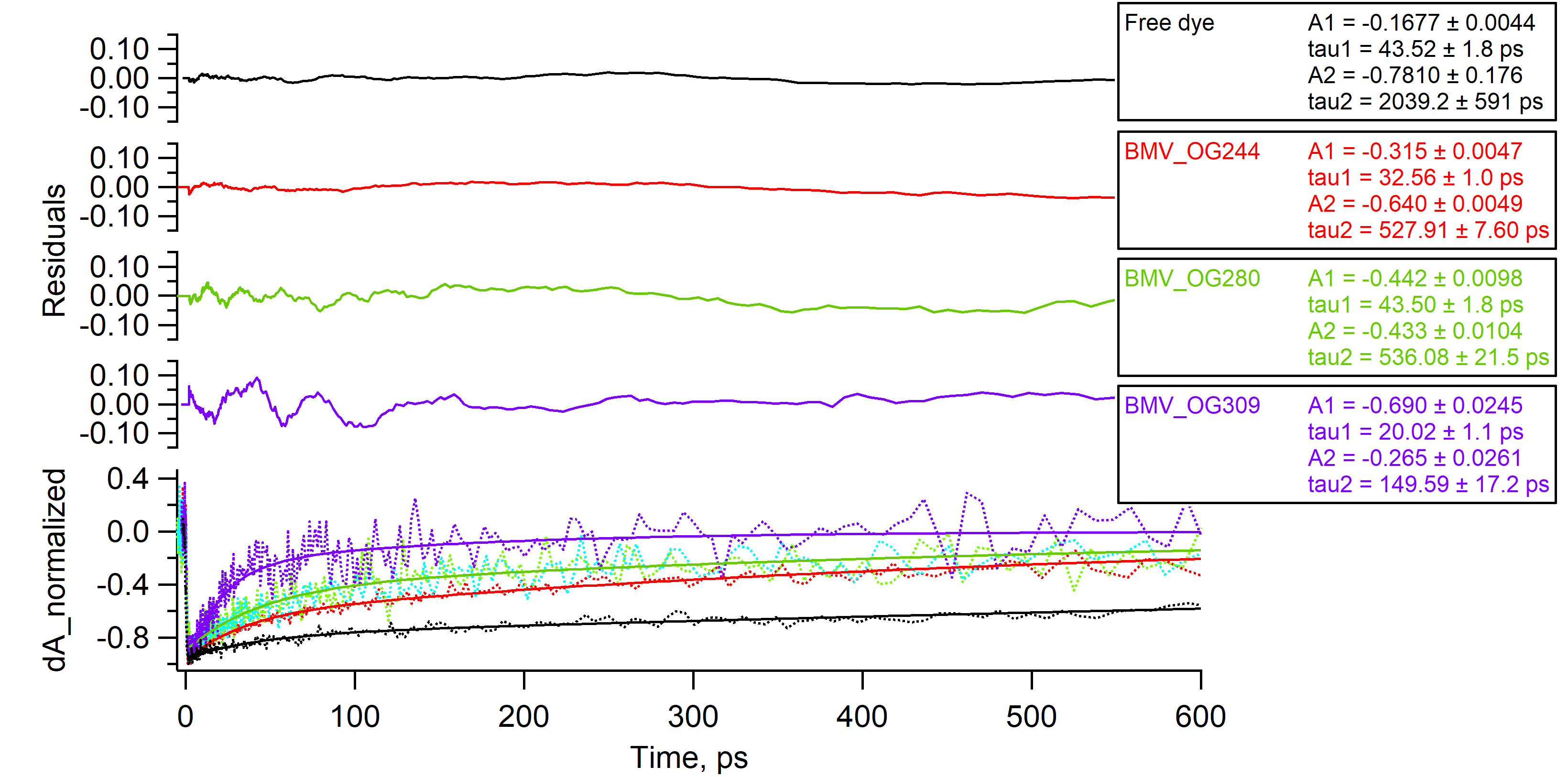}
    \caption{Oscillating signals derived from TA decay fitting. The normalized TA amplitude and fitting curves are on bottom panel of the graph and the box-filter smoothed (10 data points) residuals are on the top panels of the graph.}
    \label{ringing}
\end{figure}

The higher signal-to-noise ratio afforded by this setup change has allowed us to make an interesting observation. Figure \ref{ringing} presents normalized absorbance difference curves obtained for free dye and BMV-OG with $\braket{N}$ ranging from 200 to 300 dyes/virus. The data were tail-fitted with double exponential decay and the residuals from the fitting are presented on the top panels of the graph. At high dye loads, oscillations in TA can be observed, even in the raw data but most prominently in the residuals. 

Oscillating TA signals previously observed in photosynthetic complexes at room temperature were attributed to vibrations of the protein matrix encapsulating the fluorophores\cite{panitchayangkoon2010long,engel2011quantum}. The oscillations observed on high-load VLPs in Figure \ref{ringing} persist for a longer time, \emph{i.e.} up to 100 ps, and have lower frequency. They cannot be assigned to quantum beats from closely spaced levels since those are not expected to be observed at room temperature. We hypothesise that the oscillations have to do with the template. Two mechanisms could be envisioned by which the virus template may affect the collective emission of the multi-fluorophore antenna. First, we note that the frequency of the observed oscillations falls within the range of computed low-frequency modes (vibrational/breathing) of the virus shells \cite{vanVlijmen2005, Dykeman2008, Hadden2018}. Since coupling in near-field is very sensitive to inter-fluorophore distance and orientation, such low-frequency vibrations could modulate the TA trace. Second, the template may modulate the radiative rate of the antenna via oscillations in polarizability due to virus shell vibrations \cite{Young2018}. Could such virus-shell modes be involved in maintaining fluorophore correlations? At this point such questions concerning the role of spatial symmetry and the nature of coupling remain open, and warrant future studies.

In conclusion, the symmetric BMV capsids provides a biological scaffold for a multi-fluorophore array that acts an antenna with accelerated emission rate at room temperature. We have employed ultrafast, time-resolved transient absorption and fluorescence spectroscopy to provide further evidence for the mechanism of radiation brightening. Taken together with results of previous work, the new data strongly suggest that the virus-supported multi-fluorophore array exhibits a form of superradiant emission. Moreover, coherent, low-frequency virus shell vibrations could be observed as they modulate the relaxation rate for nearly 100 ps at room temperature. This finding highlights the involvement of the virus scaffold in the excited state relaxation dynamics. Enhanced light intensity, non-classical emission dynamics, and biocompatibility make these viromimetic superfluorescent particles interesting for biological imaging at high background conditions, for instance as deep tissue optical probes.

\newpage
\section{Experimental}
\subsection{BMV-OG sample preparation}
BMV-OG samples were prepared according to protocols previously described \cite{tsvetkova2019radiation}. Briefly, the lysines of wtBMV were modified by covalent conjugation with Oregon Green\texttrademark~488 carboxylic acid, succinimidyl ester\cite{Johnson2010}. wtBMV stock in SAMA buffer (50 mM NaOAc, 8 mM \ce{Mg(OAc)2}(pH 4.6)) was mixed with sodium bicarbonate buffer (100 mM, pH 8.2) in 1:1 ratio to a final concentration of $10^{13}$ particles/ml. Dye stock was freshly prepared in DMSO and was added to virus solution at varied ratios ranging from 500 to 10000 dyes/capsid. The DMSO concentration in reaction mixture was kept constant at $\sim5$ w/v\%. The reaction was left for 1-2 hours at room temperature, followed by removal of free dye via filter wash or dialysis with SAMA buffer. UV-Vis spectroscopy was performed to estimate the number of dyes per virus, and fluorescence spectroscopy was used to confirm absence of free dyes in the purified samples.

\subsection{Transmission Electron Microscopy}
Electron-transparent samples were prepared by placing the dilute sample (10 $\mu$L) onto a carbon-coated copper grid. After 10 mins, excess solution on the grid was removed with filter paper. The grid was stained with uranyl acetate (10 $\mu$L of 2\% solution) for 10 minutes and the excess solution was removed by blotting with filter paper. The sample was then left to dry for several minutes. Images were acquired at an accelerating voltage of 80 kV on a JEOL JEM 1010 Transmission electron microscope and analyzed with the ImageJ Processing Toolkit (see SI2).

\subsection{Time-resolved Pulsed Excitation Fluorescence Spectroscopy with Streak camera}

The sample was excited by a pulse generated through an optical parametric amplifier using a Ti:sapphire femtosecond laser (Spectra-Physics Spitfire Pro). The pulses have a wavelength of $\lambda = 480$ nm, a duration of 35 fs, and a repetition rate of 2000 Hz. The focused spot size was measured to be 160 $\mu$m and the average laser power varied from 1 to 6.5 mW (or 0.5 to 3.25 $\mu$J/pulse). The excitation beam was directed at the cuvette at a small angle to prevent cavity effect. The sample fluorescence was collected with a lens, directed through long pass filter (LP508) and a 150 mm spectrograph, then was detected as a function of wavelength and time after excitation using a photon counting streak camera (Hamamatsu C5680). Data was collected and pre-processed by HPDTA software and further analysis was done with Igor Pro software.

\subsection{Transient absorption spectroscopy}
Ultrafast transient absorption measurements were carried out using an output of regeneratively amplified Ti:sapphire laser (800 nm, 120 fs, 2 kHz repetition rate) which was split into two beams. The first beam, containing 10\%  power, was focused into a sapphire crystal to generate a white light continuum (440-750 nm), which serves as the probe laser. The other beam, containing 90\% of the power, was sent into an optical parametric amplifier to generate the 480 nm pump beam. After passing through a depolarizer, the pump beam is focused and overlapped with the probe beam at the sample (focal diameter being ~300 $\mu$m).  The pump pulse fluence was varied from 0.71 mJ/$\text{cm}^2$ to 14.5 mJ/$\text{cm}^2$, while the probe pulse fluence was significantly smaller and kept constant in all experiments. The upper and lower limits over the range of excitation energies correspond to photon densities that were close to but greater than the number of fluorophores attached in VLP antennas. Prior to any analysis, spectra were corrected for chirp and group dispersion velocity using Surface Explorer (Ultrafast Systems).

\subsection{Transient absorption set-up}
The tunable wavelength output of optical parametric amplifier pumped by a Ti:sapphire regenerative amplifier at 250 kHz repetition rate(180 fs pulsewidth), seeded by a mode-locked oscillator at 76 MHz (Coherent Inc.) was set at 480 nm and used as the pump pulse for transient absorption measurements. The part of white light output from OPA passed through a 490/20 optical filter and used as probe pulse. The fluence of the pump was varied from from 0.1 to 3.5 mJ/cm$^{2}$ and the probe pulse was set at $\sim5 ~\mu$J/mm$^{2}$. For measurements of relaxation decay, the probe pulse arrived at a delayed time after the pump that was continuously adjusted using a variable length delay line (600 ps max.). A mechanical chopper was used to modulate the pump beam at 1000 Hz. The lock-in amplifier was used to measure probe signal at the modulation frequency. The transient absorption data processing included a background subtraction step. 

\begin{acknowledgement}

The work was supported by the Army Research Office, under awards W911NF2010072 and W911NF2010071, and by the National Science Foundation, under award CBET-1803440. This work was performed, in part, at the Center for Nanoscale Materials, a U.S. Department of Energy Office of Science User Facility, and supported by the U.S. Department of Energy, Office of Science, under Contract No. DE-AC02-06CH11357. We are grateful to the Center for Bioanalytical Metrology (CBM), an NSF Industry-University Cooperative Research Center, for providing funding for this project under grant NSF IIP 1916645, and to members of the industry advisory board of the CBM for valuable discussions and feedback.

\end{acknowledgement}

\begin{suppinfo}

The Supporting Information is available free of charge on the ACS Publications website at DOI: 10.1021/acs.jpclett.0000000. Experimental methods, sample characterization, TA fitting results and model of superradiant emission from N fluorophores are presented in the supporting information file.

\end{suppinfo}

\bibliography{ANL_Streak_TA}

\end{document}


\newpage
\subsection{Samples characterization}
\begin{figure}[H]
    \centering
    \includegraphics[width = 0.8 \textwidth]{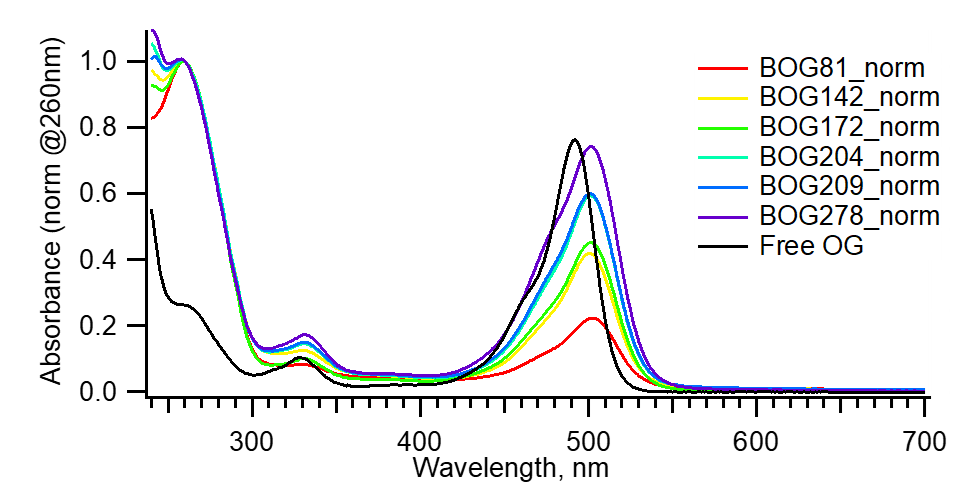}
    \caption{The average number of dyes in the labelled virus-like particles was obtained using UV-Visible absorption spectroscopy\cite{tsvetkova2019radiation}. The average dye numbers were estimated using a molar extinction coefficient value of $\varepsilon$ = 33,641 cm$^{-1}$M$^{-1}$ at pH 4.6 for the Oregon green NHS ester. The absorbance of free dye is shown in black for comparison. The legends depict samples with varying $<N>$ - number of dyes per virus. The spectra of OG conjugated BMV are normalized at 260 nm.}
    \label{UV-Vis}
\end{figure}

\begin{figure}[H]
    \centering
    \includegraphics[width = 0.5 \textwidth]{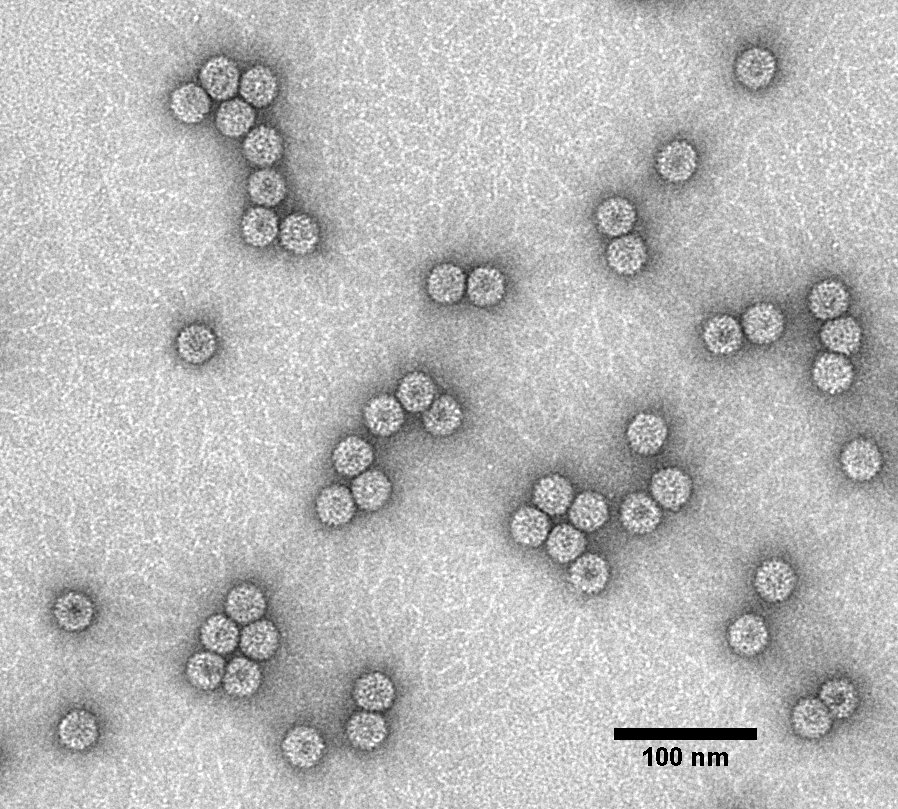}
    \caption{The BMV-OG particles morphology was checked by TEM to confirm they are identical to wt BMV. TEM image shown for BMV-OG with 209 dyes. All samples were checked before and after spectroscopy experiments.}
    \label{TEM}
\end{figure}

\newpage
\subsection{Spectral Evolution of Fluorophore Relaxation}

Evolution of the difference spectra from $900$fs to $2600$ps are shown in figures SI3-7. Femtosecond-TA data in this set of plots were pre-processed following a penalized least squares method that involved a third order penalty\cite{Eilers2003}. We note that this method has been used previously by other groups to remove artifact contributions and as a baseline correction method in the treatment of ultrafast spectrokinetic data \cite{Stokkum2004,Devos2011,Ruckebusch2012}. The break in curves is due to removal of pump laser scattering. 

The major spectral features observed in the evolution-associated difference spectra (SI3-7) are a large negative region between $500-550$ nm, which is flanked on either side by two positive regions in the $425-470$ nm and $700-725$ nm ranges. The ground state population is quickly depleted following application of the pump pulse and results in a strong negative signal. Therefore, the negative signal at $500-550$ nm can easily be attributed to the GSB of the fluorophore. Positive features are indicative of excited state absorption (ESA) in the areas on either side of the GSB signal. In each of the following Figures SI3-7, the spectral evolution are plotted at three pump fluences of $0.71 \text{mJ}/\text{cm}^2$ (A), $7.1 \text{mJ}/\text{cm}^2$ (B), and $14.5 \text{mJ}/\text{cm}^2$ (C), while what is changing between figures is the $\braket{N}$.

\begin{figure}[H]
    \centering
    \includegraphics[width=\textwidth]{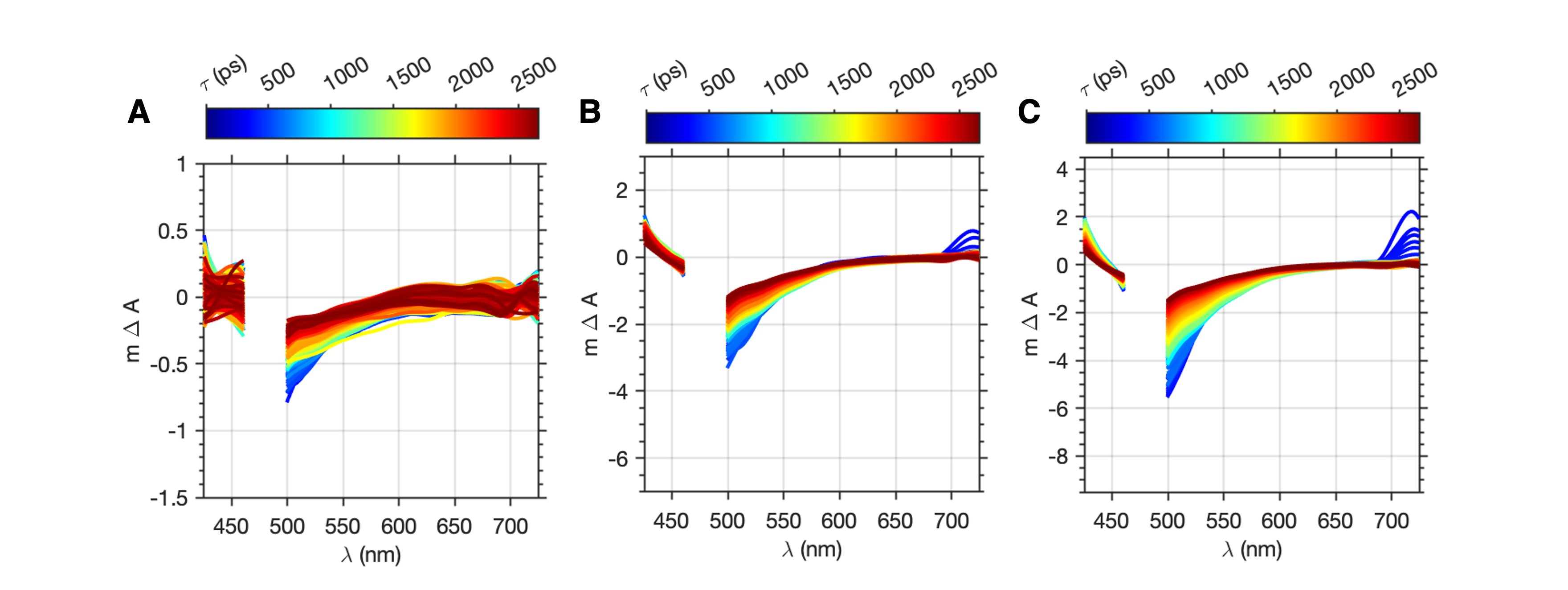}
    \caption{Spectral evolution of free OG fluorophore in solution, $\braket{N}=0$. Pump fluences are $0.71 \text{mJ}/\text{cm}^2$ (A), $7.1 \text{mJ}/\text{cm}^2$ (B), and $14.5 \text{mJ}/\text{cm}^2$ (C).}
    \label{OG_EADS}
\end{figure}

\begin{figure}[H]
    \centering
    \includegraphics[width=\textwidth]{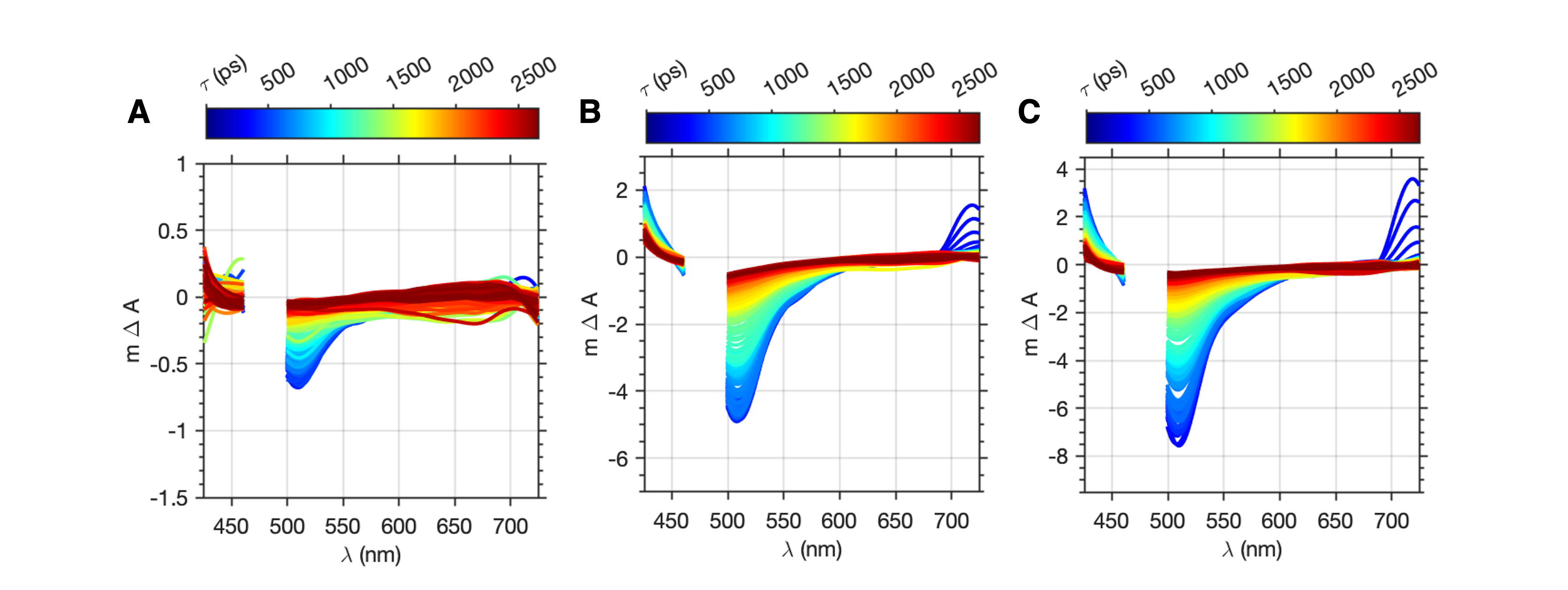}
    \caption{Spectral evolution of a BMV-OG multifluorophore antenna with $\braket{N}=81$. Pump fluences are $0.71 \text{mJ}/\text{cm}^2$ (A), $7.1 \text{mJ}/\text{cm}^2$ (B), and $14.5 \text{mJ}/\text{cm}^2$ (C).}
    \label{BOG81_EADS}
\end{figure}

\begin{figure}[H]
    \centering
    \includegraphics[width=\textwidth]{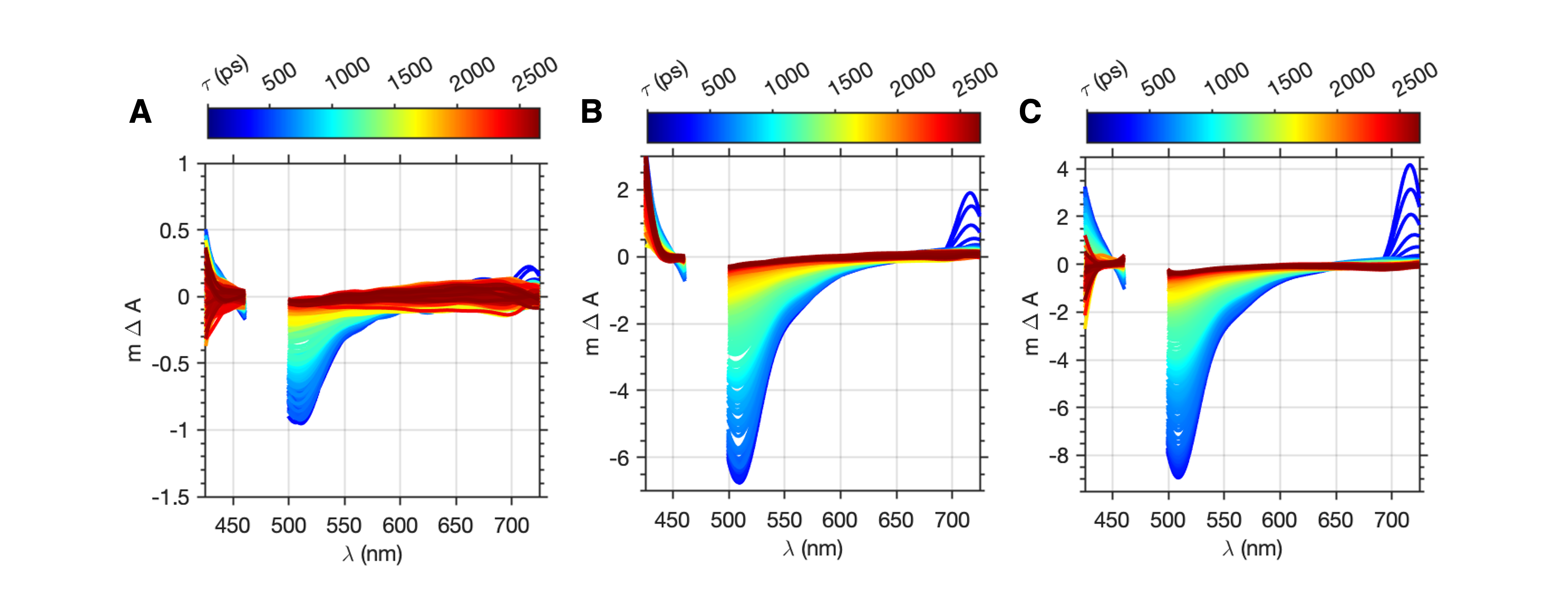}
    \caption{Spectral evolution of a BMV-OG multifluorophore antenna with $\braket{N}=172$. Pump fluences are $0.71 \text{mJ}/\text{cm}^2$ (A), $7.1 \text{mJ}/\text{cm}^2$ (B), and $14.5 \text{mJ}/\text{cm}^2$ (C).}
    \label{BOG172_EADS}
\end{figure}

\begin{figure}[H]
    \centering
    \includegraphics[width=\textwidth]{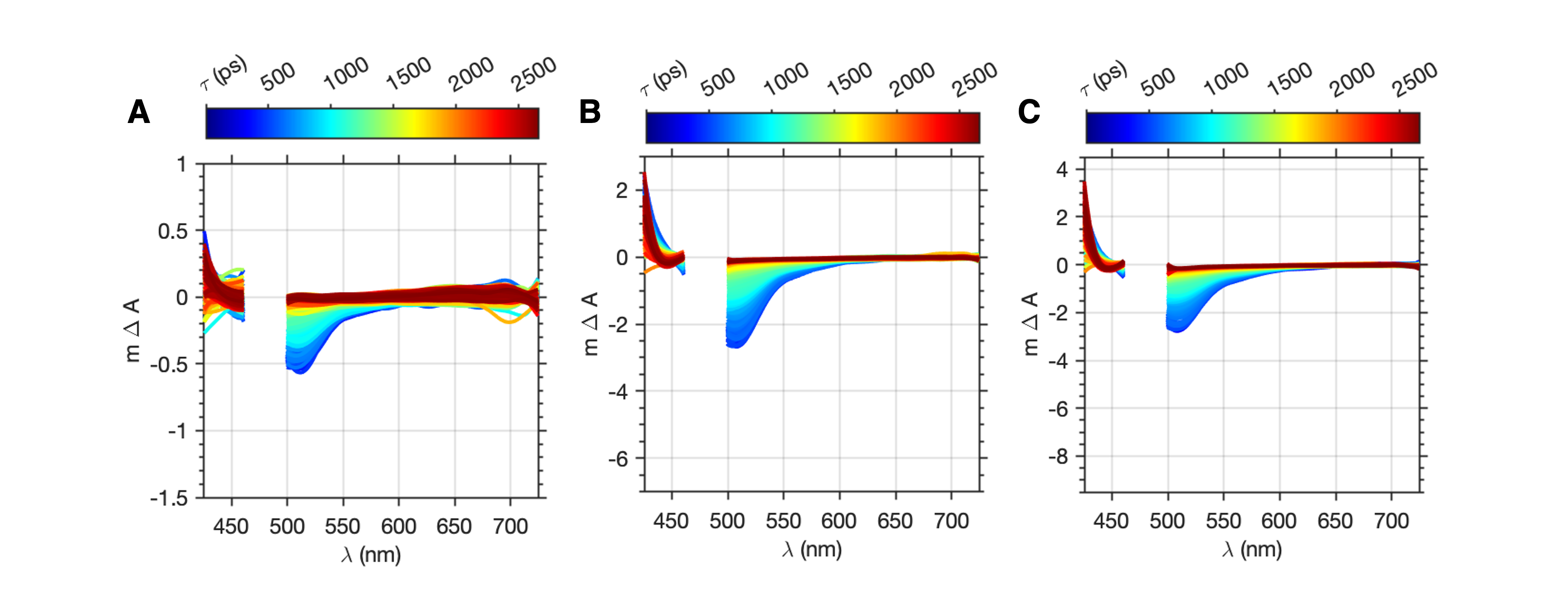}
    \caption{Spectral evolution of a BMV-OG multifluorophore antenna with $\braket{N}=209$. Pump fluences are $0.71 \text{mJ}/\text{cm}^2$ (A), $7.1 \text{mJ}/\text{cm}^2$ (B), and $14.5 \text{mJ}/\text{cm}^2$ (C).}
    \label{BOG209_EADS}
\end{figure}

\begin{figure}[H]
    \centering
    \includegraphics[width=\textwidth]{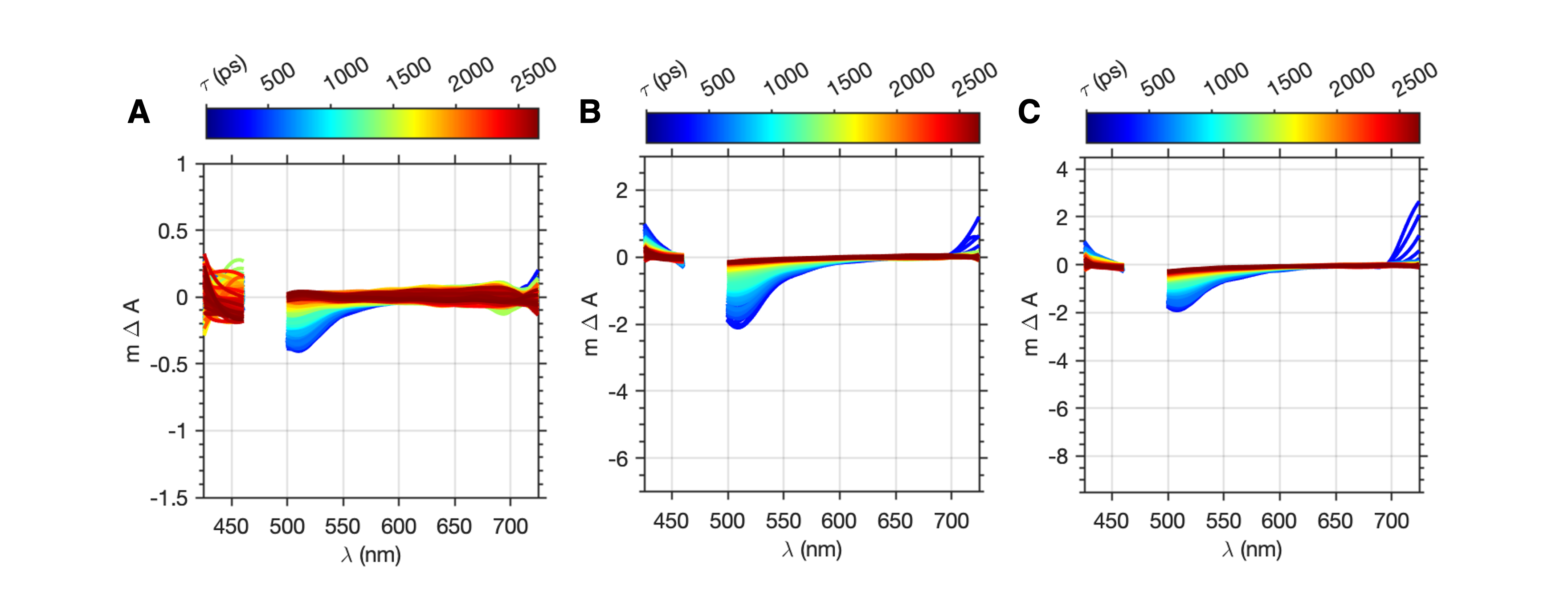}
    \caption{Spectral evolution of a BMV-OG multifluorophore antenna with $\braket{N}=278$. Pump fluences are $0.71 \text{mJ}/\text{cm}^2$ (A), $7.1 \text{mJ}/\text{cm}^2$ (B), and $14.5 \text{mJ}/\text{cm}^2$ (C).}
    \label{BOG278_EADS}
\end{figure}

\subsection{Kinetic Models of the GSB decay }
To examine the time-dependent GSB decay, we integrated the unprocessed difference spectra obtained by femtosecond-TA over the $520-530$ nm region to extract a single kinetic trace at the central wavelength of $525$ nm. Due to pump-pulse related artifacts, which are observed at very early times, the spectrokinetic data were fit from 400 fs to 2.6 ns using an exponential model.  Each exponential term was given an individual amplitude $A_i$ and a characteristic time constant $\tau_i$ -- lifetime:
\begin{equation}
    S(t)= \sum_{i} A_i e^{-(t-t_0)/\tau_i} .
\end{equation}
Following the fitting procedure, individual amplitudes were subsequently normalized by the sum of $A_i$'s to reveal their overall fractional contribution to the recovery kinetics.

A single exponential decay adequately describes the features present in the recovery curves of a single fluorophore, while two terms were necessary for the case of bound fluorophore in VLP antenna systems. In each of the following figures (SI8-12), model fit functions (red curve) are plotted against the integrated decay trace (open black circles) and the corresponding residuals associated with the model fit are plotted in the top panel at each pump fluence of $0.71 \text{mJ}/\text{cm}^2$ (A), $7.1 \text{mJ}/\text{cm}^2$ (B), and $14.5 \text{mJ}/\text{cm}^2$ (C), respectively. The different figures are established by their different $\braket{N}$ in figures SI8-12.

\begin{figure}[H]
    \centering
    \includegraphics[width=\textwidth]{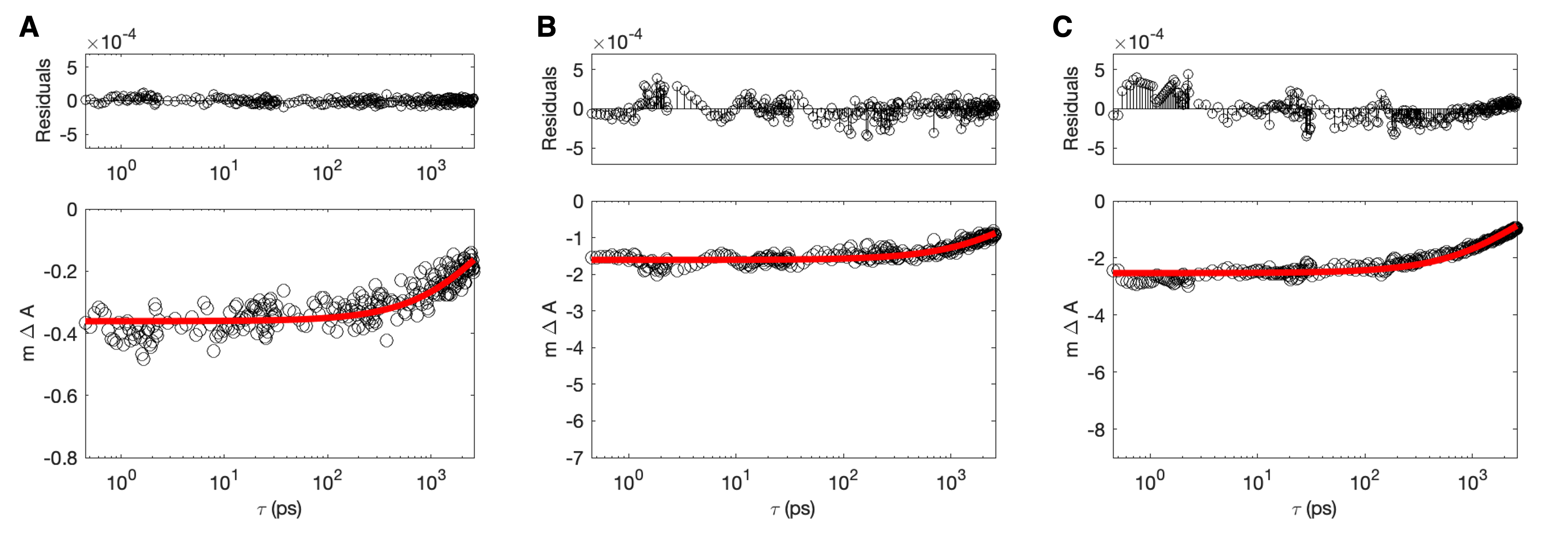}
    \caption{Kinetic models of GSB decay for free OG fluorophore in solution, $\braket{N}=0$. Pump fluences are $0.71 \text{mJ}/\text{cm}^2$ (A), $7.1 \text{mJ}/\text{cm}^2$ (B), and $14.5 \text{mJ}/\text{cm}^2$ (C).}
    \label{Fits_N=0}
\end{figure}

\begin{figure}[H]
    \centering
    \includegraphics[width=\textwidth]{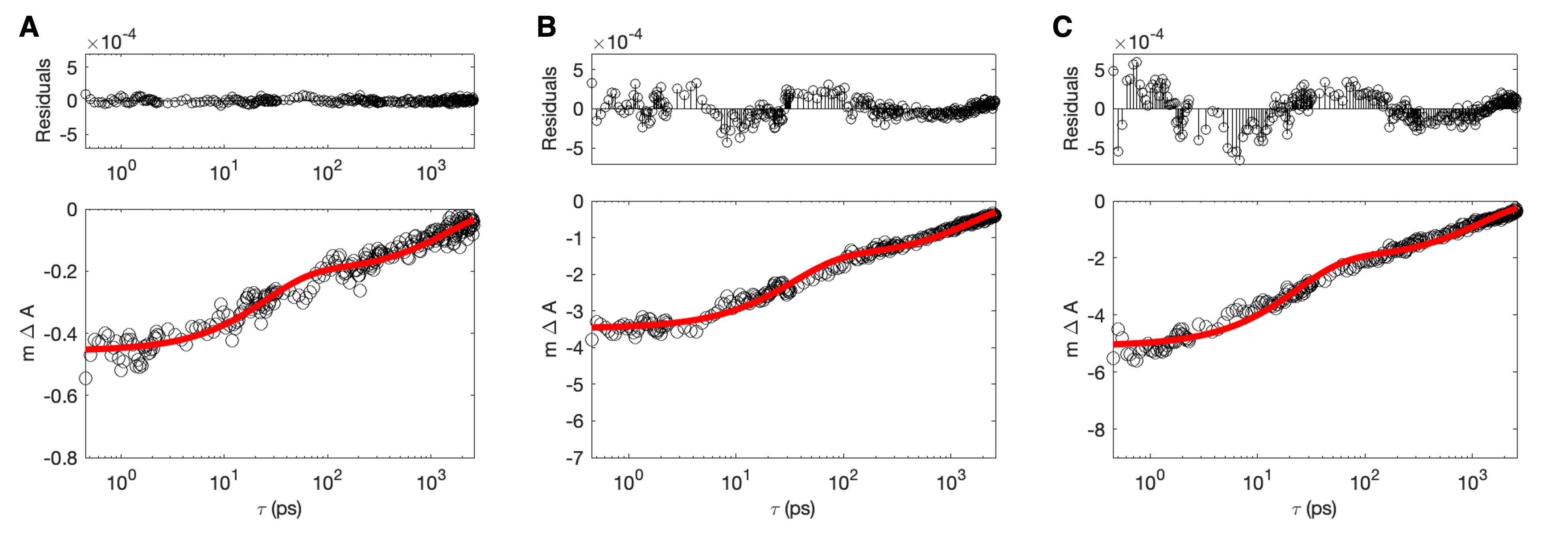}
    \caption{Kinetic models of GSB decay for a BMV-OG multifluorophore with $\braket{N}=81$. Pump fluences are $0.71 \text{mJ}/\text{cm}^2$ (A), $7.1 \text{mJ}/\text{cm}^2$ (B), and $14.5 \text{mJ}/\text{cm}^2$ (C).}
    \label{Fits_N=81}
\end{figure}

\begin{figure}[H]
    \centering
    \includegraphics[width=\textwidth]{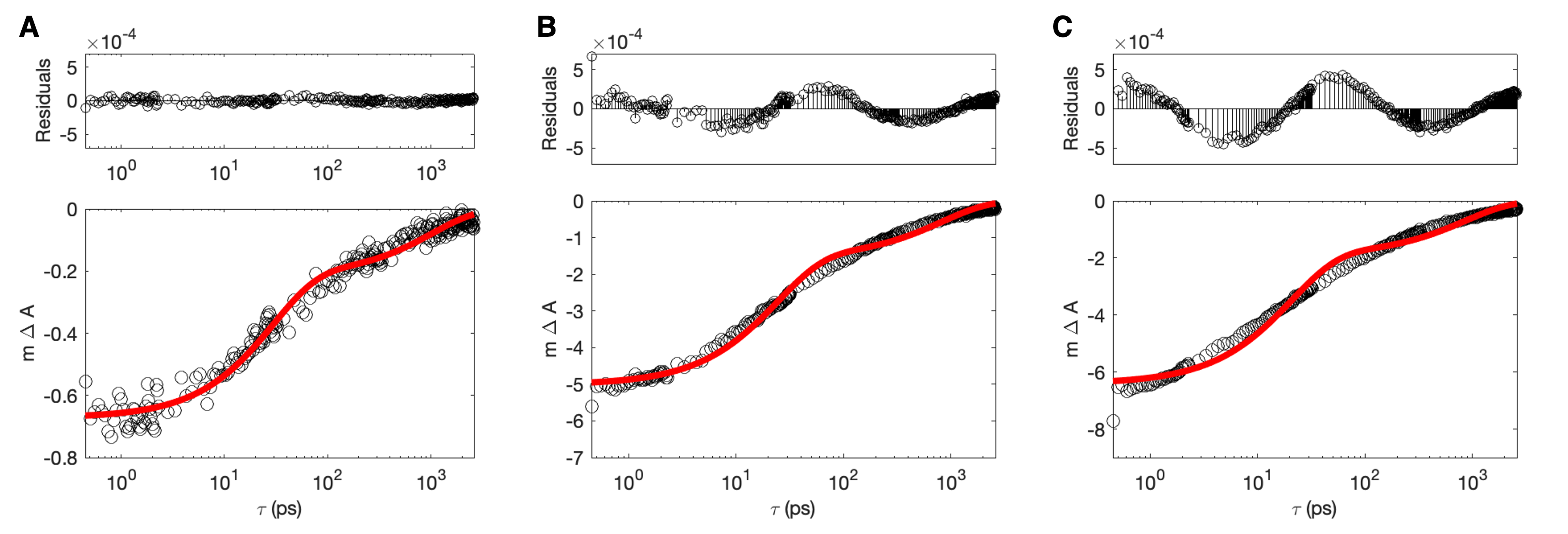}
    \caption{Kinetic models of GSB decay for a BMV-OG multifluorophore with $\braket{N}=172$. Pump fluences are $0.71 \text{mJ}/\text{cm}^2$ (A), $7.1 \text{mJ}/\text{cm}^2$ (B), and $14.5 \text{mJ}/\text{cm}^2$ (C).}
    \label{Fits_N=172}
\end{figure}

\begin{figure}[H]
    \centering
    \includegraphics[width=\textwidth]{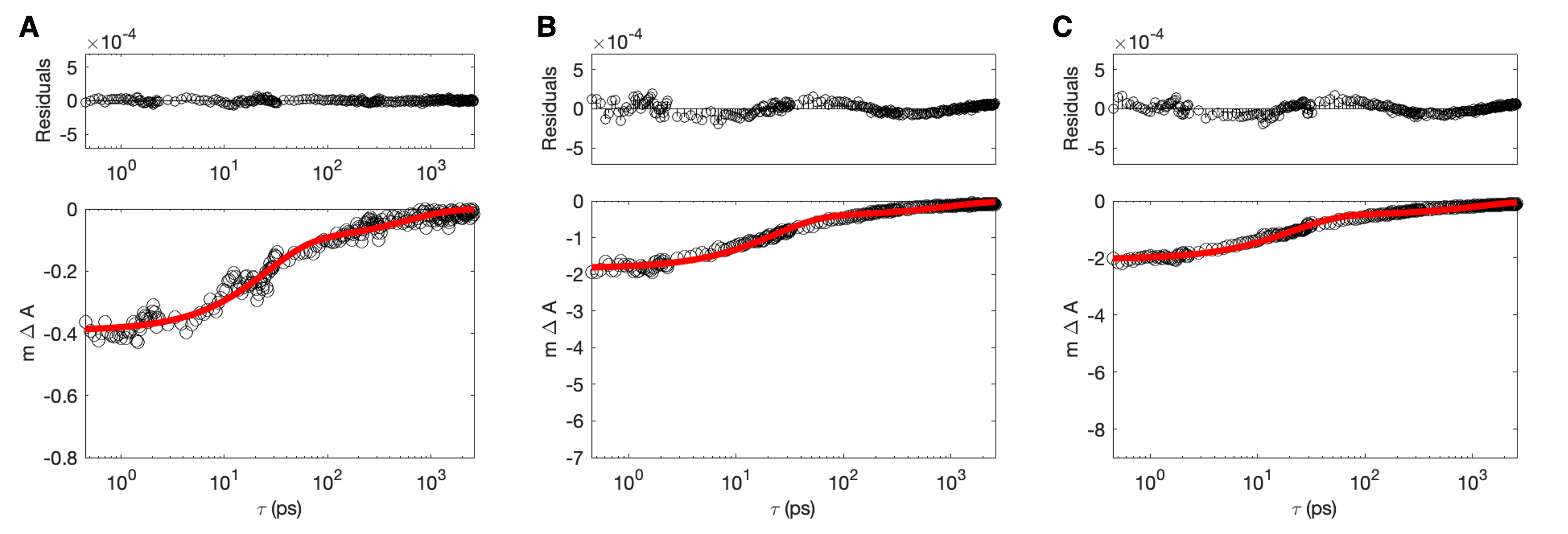}
    \caption{Kinetic models of GSB decay for a BMV-OG multifluorophore with $\braket{N}=209$. Pump fluences are $0.71 \text{mJ}/\text{cm}^2$ (A), $7.1 \text{mJ}/\text{cm}^2$ (B), and $14.5 \text{mJ}/\text{cm}^2$ (C).}
    \label{Fits_N=209}
\end{figure}

\begin{figure}[H]
    \centering
    \includegraphics[width=\textwidth]{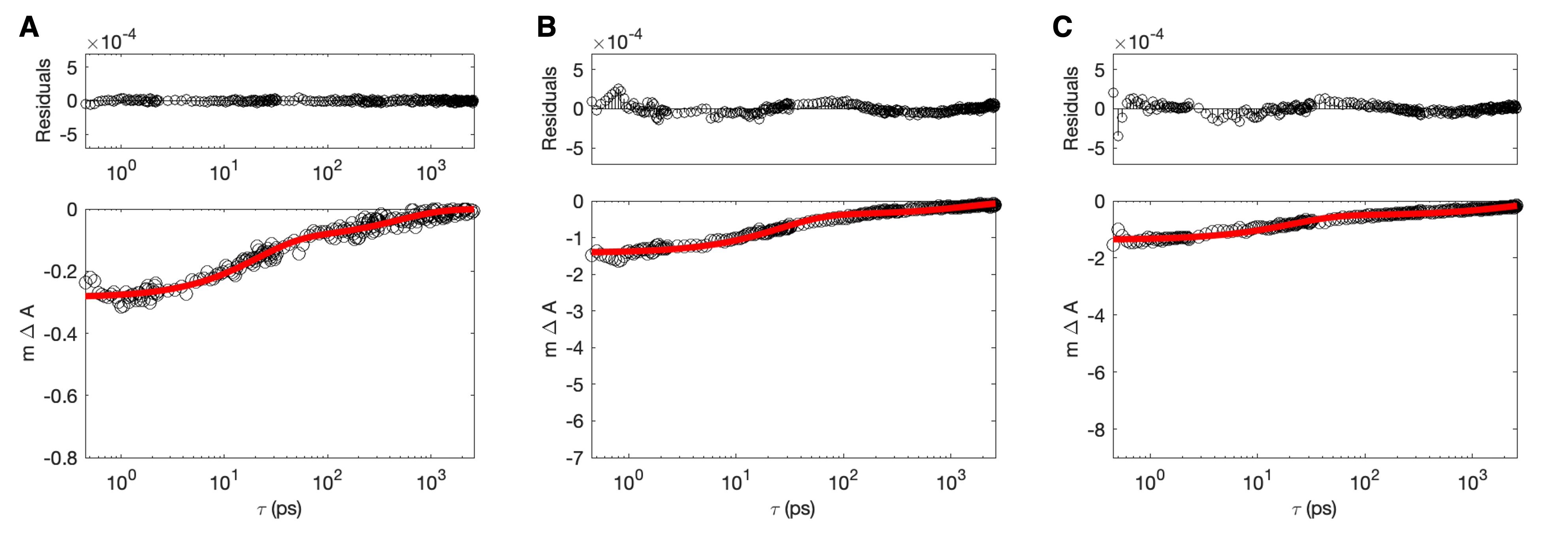}
    \caption{Kinetic models of GSB decay for a BMV-OG multifluorophore with $\braket{N}=278$. Pump fluences are $0.71 \text{mJ}/\text{cm}^2$ (A), $7.1 \text{mJ}/\text{cm}^2$ (B), and $14.5 \text{mJ}/\text{cm}^2$ (C).}
    \label{Fits_N=278}
\end{figure}

\subsection*{A simple model of superradiant emission dynamics}

Consider a collection of N identical 2-level entities, each with a ground state, $\ket{g}$, and an excited state, $\ket{e}$. We assume they are fixed rigidly in place (to avoid recoil effects) and are much closer to each other than the wavelength of the radiation they can emit, but not so close that direct interactions (e.g. Van der Waals couplings) become significant. The initial condition at time $t=0$s that all the entities are in the excited state, a symmetric arrangement. Here we assume that the spontaneous emission process is much more rapid than nonradiative decays and neglect the latter. Each entity is dipole coupled in the same way to the electromagnetic field so the perturbation term in the Hamiltonian is symmetric in its action.

As Dicke first described\cite{Dicke1954},  the system should spread down a “ladder” of purely symmetric states. We write for each rung of the ladder the state $\ket{N,s}$ where $0 \leq s \leq N$ is the number of entities in the ground state at that rung. One does not need to specify which entities are in their ground state because $\ket{N,s}$  formally includes all possible ways of distributing s ground states among the $N$ entities. These collective states are readily orthonormalized. The time dependent wavefunction for the collection of entities may be written as
\begin{equation}
    \ket{N,t}=\sum_{s} \alpha_s(t) \ket{N,s}
\end{equation}
and the focus is on the $N+1$ occupational probabilities $\Pi_s(t) = |\alpha_s(t)|^2$, which satisfy, $\sum_{s} \Pi_s(t) = 1$. The initial condition has $\Pi_s(0)=\delta_{s,0}$ and the system decays towards $\Pi_s(\infty) = \delta_{s,N}$ with N photons emitted. For intermediate times one needs to solve the coupled equations\cite{Gross1982}
\begin{equation}
    \frac{\partial {\Pi_s}}{\partial t} = -\Gamma_s\Pi_s + \Gamma_{s-1}\Pi_{s-1},
\end{equation}
which describe how $\Pi_s$ changes due to decays to $\ket{N,s+1}$ and gains from $\ket{N,s-1}$. The rate factors, $\Gamma_s$, are given by
\begin{equation}
    \Gamma_s = (N-s)(s+1) \Gamma , 
\end{equation}
where $\Gamma$ is the decay rate of a single, isolated entity. Note that for $s$ near $N/2$, $\Gamma_s/N\Gamma \approx N/4$. The rate of photon emission is $W(t) = \sum_{s} \Gamma_s \Pi_s(t)$. 

\begin{figure}[H]
    \centering
    \includegraphics[width=\textwidth]{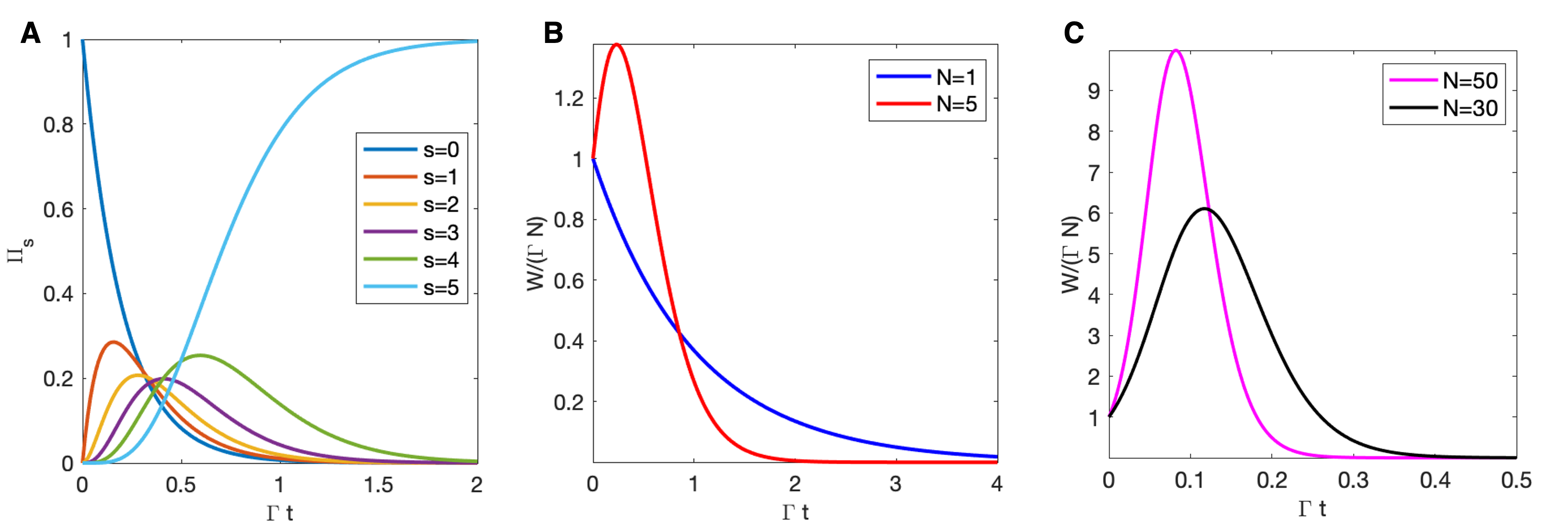}
    \caption{(A) Occupational probabilities as a function of the normalized time for $N=5$ identical molecules. (B) Photon emission rates for $N=5$ independent (blue) and coupled (red) molecules. (C) Emission dynamics for the case of $N=30$ (black) and $N=50$ (magenta) molecules.}
    \label{Rate}
\end{figure}

Figure \ref{Rate} (A) shows the evolution/spreading of the occupational probabilities for the case of $N=5$ coupled entities. The photon emission rate for $N=5$ collective (red curve) and that of independent emitters (blue curve) is given in Figure \ref{Rate} (B). Previous work by Anil \emph{et al.}\cite{AnilSushma2021} identified only a subset of the total number of fluorophores attached in VLP antennas are responsible for the observed radiation brightening phenomena\cite{AnilSushma2021}. Therefore, we compare an "effective" N of $N=30$ (black) or $N=50$ (magenta) coupled fluorophores in our calculation of the photon emission rates (Figure \ref{Rate}(C)). Both the numerical calculations and experimental results show a delay or lag time before the peak in the emitted intensity is reached, however, the delay is considerably shorter in the experiment (by an order of magnitude). This just might be due to the simplicity of our theory which does not take into account the effect of non-radiative relaxation. Thus, other short/quick relaxation pathways may compete with the collective one leading to a shortening of the delay time and/or decreasing the peak amplitude for the case of virus multifluorophore antennas at room temperature.

\bibliography{ANL_SI}